\documentclass[11pt]{article}

\usepackage{mathpazo}
\usepackage{amsfonts}
\usepackage{amsmath}
\usepackage{amssymb}
\usepackage{latexsym}
\usepackage{mathtools}

\usepackage[margin=1in]{geometry}
\usepackage{hyperref}
\hypersetup{pdfpagemode=UseNone}
\usepackage{amsthm}

\theoremstyle{definition}

\usepackage{authblk}
\usepackage{tikz}
\usetikzlibrary{calc,positioning}
\usepackage{mdframed}

\mdfdefinestyle{figstyle}{
  linecolor=black!7,
  backgroundcolor=black!7,
  innertopmargin=10pt,
  innerleftmargin=25pt,
  innerrightmargin=25pt,
  innerbottommargin=10pt
}

\definecolor{White}{rgb}{1,1,1}
\definecolor{Black}{rgb}{0,0,0}
\definecolor{LightGray}{rgb}{.81,.81,.81}
\colorlet{ChannelColor}{LightGray}
\colorlet{ChannelTextColor}{Black}
\colorlet{ReadoutColor}{White}

\newcommand{\microspace}{\mspace{0.5mu}}

\newcommand{\tr}{\operatorname{Tr}}

\renewcommand{\int}{\operatorname{int}}

\renewcommand{\t}{{\scriptscriptstyle\mathsf{T}}}

\newcommand{\ip}[2]{\langle #1 , #2\rangle}
\newcommand{\bigip}[2]{\bigl\langle #1, #2 \bigr\rangle}
\newcommand{\Bigip}[2]{\Bigl\langle #1, #2 \Bigr\rangle}

\newcommand{\ket}[1]{
  \lvert\microspace #1 \microspace \rangle}

\newcommand{\bra}[1]{
  \langle\microspace #1 \microspace \rvert}

\newcommand{\I}{\mathbb{1}}

\newcommand{\setft}[1]{\mathrm{#1}}
\newcommand{\Density}{\setft{D}}
\newcommand{\Pos}{\setft{Pos}}
\newcommand{\Unitary}{\setft{U}}

\newcommand{\Lin}{\setft{L}}

\newcommand{\complex}{\mathbb{C}}

\renewcommand{\natural}{\mathbb{N}}

\newenvironment{mylist}[1]{\begin{list}{}{
	\setlength{\leftmargin}{#1}
	\setlength{\rightmargin}{0mm}
	\setlength{\labelsep}{2mm}
	\setlength{\labelwidth}{8mm}
	\setlength{\itemsep}{0mm}}}
	{\end{list}}

\newcommand{\X}{\mathcal{X}}
\newcommand{\Y}{\mathcal{Y}}

\newcommand{\V}{\mathcal{V}}
\newcommand{\U}{\mathcal{U}}

\newcommand{\R}{\mathcal{R}}

\renewcommand{\S}{\mathcal{S}}

\newcommand{\reg}[1]{\textsf{#1}}

\makeatletter
\let\@fnsymbol\@arabic
\makeatother

\begin{document}

\title{\LARGE\bf Extended Nonlocal Games from Quantum-Classical Games}

\author[1]{Vincent Russo}
\author[1,2]{John Watrous}

\affil[1]{Institute for Quantum Computing and School of Computer
  Science\protect\\
  University of Waterloo, Canada\vspace{2mm}}
  
\affil[2]{Canadian Institute for Advanced Research\protect\\
  Toronto, Canada}

\date{September 6, 2017}

\renewcommand\Affilfont{\normalsize\itshape}
\renewcommand\Authfont{\large}
\setlength{\affilsep}{6mm}
\renewcommand\Authand{\rule{10mm}{0mm}}

\maketitle

\begin{abstract}
  Several variants of nonlocal games have been considered in the study of
  quantum entanglement and nonlocality.
  This paper concerns two of these variants, called 
  \emph{quantum-classical games} and \emph{extended nonlocal games}.
  We give a construction of an extended nonlocal game from any
  quantum-classical game that allows one to translate certain facts concerning
  quantum-classical games to extended nonlocal games.
  In particular, based on work of Regev and Vidick, we conclude that there
  exist extended nonlocal games for which no finite-dimensional entangled
  strategy can be optimal.
  While this conclusion is a direct consequence of recent work of Slofstra,
  who proved a stronger, analogous result for ordinary (non-extended) nonlocal
  games, the proof based on our construction is considerably simpler, and the
  construction itself might potentially have other applications in the study of
  entanglement and nonlocality.
\end{abstract}

%------------------------------------------------------------------------------%
\section{Introduction}
\label{sec:intro}
%------------------------------------------------------------------------------%

Various abstract notions of \emph{games} have been considered in the study of
entanglement and nonlocality
\cite{CleveHTW04, BrassardBT05, CleveSUU08, DohertyLTW08, KempeKMTV08,
  KempeRT08, Buscemi12, Fritz12, LeungTW13, TomamichelFKW13, CleveM14,
  CooneyJPP15, RegevV15, JohnstonMRW16}.
For instance, in a \emph{nonlocal game}, two cooperating players (Alice and
Bob) engage in an interaction with a third party (known as the \emph{referee})
\cite{CleveHTW04}.
The referee randomly chooses a pair of questions $(x,y)$ according to a known
distribution.
Alice receives $x$, Bob receives $y$, and without communicating with one
another, Alice must respond with an answer $a$ and Bob with an answer~$b$.
The referee then evaluates a predicate $P(a,b|x,y)$ to determine whether Alice
and Bob win or lose.
It is a well-known consequence of earlier work in theoretical physics
\cite{Bell64,KochenS67,ClauserHSH69} that entanglement shared between Alice and
Bob can allow them to outperform all purely classical strategies for some
nonlocal games.
(Nonlocal games were also previously studied in theoretical computer science,
in \cite{Raz98} for instance, although generally not by this name and
without deference to entanglement or quantum information, but rather as an
abstraction of one-round, two-player classical interactive proof systems.)

In a nonlocal game, the referee is classical; it is only the players Alice and
Bob that potentially manipulate quantum information.
Some generalizations of nonlocal games in which quantum information is exchanged
in some way between the players and the referee include ones studied in
\cite{Buscemi12, Fritz12, LeungTW13, TomamichelFKW13, CooneyJPP15, RegevV15,
  JohnstonMRW16}.
In this paper we consider two such generalizations:
\emph{quantum-classical games} and \emph{extended nonlocal games}.
\begin{mylist}{8mm}
\item[1.] \emph{Quantum-classical games}.
  Quantum-classical games, or \emph{QC games} for short, differ from nonlocal
  games in that the referee begins the game by preparing a tripartite quantum
  state, then sends one part of it to each player and keeps the third part for
  itself.
  (This step replaces the generation of a classical question pair $(x,y)$ in an
  ordinary nonlocal game.)
  The players respond with classical answers $a$ and $b$ as before, and finally
  the referee determines whether the players win or lose by measuring its part
  of the original quantum state it initially prepared.
  (This step replaces the evaluation of a predicate $P(a,b|x,y)$ in an ordinary
  nonlocal game.)

  Games of this form, with slight variations from the general class just
  described, were considered by Buscemi \cite{Buscemi12} and
  Regev and Vidick \cite{RegevV15}.

\item[2.] \emph{Extended nonlocal games}. 
  In an extended nonlocal game, Alice and Bob first present the referee with a
  quantum system of a fixed size, initialized as Alice and Bob choose, and
  possibly entangled with systems held by Alice and Bob.
  (This initialization step generalizes the sharing of entanglement
  between Alice and Bob in an ordinary nonlocal game, allowing them to give a
  part of this shared state to the referee.)
  The game then proceeds much like an ordinary nonlocal game:
  the referee chooses a pair of (classical) questions $(x,y)$ according to
  a known distribution, sends $x$ to Alice and $y$ to Bob, and receives
  a classical answer $a$ from Alice and $b$ from Bob.
  Finally, to determine whether or not Alice and Bob win, the referee performs
  a binary-valued measurement, depending on $x$, $y$, $a$, and $b$,
  on the system initially sent to it by Alice and Bob.
  (This measurement replaces the evaluation of the predicate $P(a,b|x,y)$ in an
  ordinary nonlocal game.)

  Games of this form, again with a slight variation from the general class just
  described, were considered by Fritz \cite{Fritz12}, who called them
  \emph{bipartite steering games}.
  Extended nonlocal games represent a game-based formulation of the phenomenon
  of \emph{tripartite steering} investigated in
  \cite{CavalcantiSANRW15,SainzBCSV15}.
  (The clash in nomenclature reflects one's view of the referee's role either
  as a non-player in a game or as a participant in an experiment.)
  Extended nonlocal games were so-named and studied in \cite{JohnstonMRW16}, as
  a means to unify nonlocal games with the 
  \emph{monogamy-of-entanglement games} introduced in \cite{TomamichelFKW13}.
\end{mylist}

Regev and Vidick \cite{RegevV15} proved that certain QC games have the
following peculiar property: if Alice and Bob make use of an entangled state of
two finite-dimensional quantum systems, initially shared between them, they can
never achieve perfect optimality: it is always possible for them to do better
(meaning that they win with a strictly larger probability) using some different
shared entangled state on two larger quantum systems.
Thus, it is only in the limit, as the local dimensions of their shared entangled
states goes to infinity, that they can approach an optimal performance in these
specific examples of games.
A similar result was established earlier for analogues of nonlocal games for
which both the questions and answers are quantum \cite{LeungTW13},
and a recent breakthrough result of Slofstra \cite{Slofstra17} has established
a similar result for nonlocal games in which both the questions and answers are
classical.

In this paper we describe a construction through which any QC game can be
transformed into an extended nonlocal game, in such a way that basic properties
associated with entangled strategies for the QC game are inherited by the
extended nonlocal game.
In particular, by applying this construction to the QC games identified by
Regev and Vidick, we obtain extended nonlocal games that cannot be played
with perfect optimality by Alice and Bob using an entangled state on
finite-dimensional systems.
In the language of quantum steering, this yields a tripartite steering
inequality for which a maximal violation requires infinite-dimensional quantum
systems.
While Slofstra's result subsumes this result, insofar as nonlocal games are
special cases of extended nonlocal games in which the referee's quantum
system is a trivial one-dimensional system, our proof is considerably simpler.
Moreover, this ability to transform from quantum-classical games to extended
nonlocal games might potentially find utility in related settings.

%------------------------------------------------------------------------------%
\section{Definitions}
\label{sec:preliminaries}
%------------------------------------------------------------------------------%

We begin with precise definitions of the two classes of games considered in
this paper, which are QC games and extended nonlocal games.
In addition, we formalize the notions of \emph{entangled strategies} for these
games along with their associated \emph{values}, which represent the
probabilities that the strategies lead to a win for Alice and Bob.

The reader is assumed to be familiar with standard notions of quantum
information, as described in \cite{NielsenC00} and \cite{Wilde13}, for
instance.
We will generally follow the terminology and notational conventions of
\cite{Watrous16}.
For example, a \emph{register} $\reg{X}$ is an abstract quantum system
described by a finite-dimensional complex Hilbert space $\X$ having a fixed
standard basis $\{\ket{1},\ldots,\ket{n}\}$ (for some positive integer $n$);
the sets $\Lin(\X)$, $\Pos(\X)$, $\Density(\X)$, and $\Unitary(\X)$ denote the
set of all linear operators, positive semidefinite operators, density
operators, and unitary operators (respectively) acting on such a space $\X$;
we write $X^{\ast}$, $\overline{X}$, and $X^{\t}$ to refer to the
adjoint, entry-wise complex conjugate, and transpose of an operator $X$
(with respect to the standard basis in the case of the entry-wise complex
conjugate and transpose); and $\ip{X}{Y} = \tr(X^{\ast} Y)$ denotes the
Hilbert-Schmidt inner product of operators $X$ and $Y$.

\subsection{Extended nonlocal games}

An \emph{extended nonlocal game} is specified by the following objects:
\begin{mylist}{8mm}
\item[$\bullet$]
  A probability distribution $\pi:X\times Y \rightarrow [0,1]$, for finite and
  nonempty sets $X$ and $Y$.
\item[$\bullet$]
  A collection of measurement operators 
  $\{P_{a,b,x,y}\,:\,a\in A,\; b\in B,\; x\in X,\; y\in Y\} \subset \Pos(\R)$,
  where $A$ and $B$ are finite and nonempty sets and $\R$ is the space
  corresponding to a register $\reg{R}$.
\end{mylist}
From the referee's perspective, such a game is played as follows:
\begin{mylist}{8mm}
\item[1.]
  Alice and Bob present the referee with the register $\reg{R}$, which has been
  initialized in a state of Alice and Bob's choosing.
  (The register $\reg{R}$ might, for instance, be entangled with systems
  possessed by Alice and Bob.)
\item[2.]
  The referee randomly generates a pair $(x,y) \in X\times Y$ according to the
  distribution $\pi$, and then sends $x$ to Alice and $y$ to Bob.
  Alice responds with $a\in A$ and Bob responds with $b\in B$.
\item[3.]
  The referee measures $\reg{R}$ with respect to the binary-valued
  measurement $\{P_{a,b,x,y},\, \I-P_{a,b,x,y}\}$.
  The outcome corresponding to the measurement operator $P_{a,b,x,y}$ indicates
  that Alice and Bob \emph{win}, while the other measurement result indicates
  that they \emph{lose}.
\end{mylist}

There are various classes of \emph{strategies} that may be considered for
Alice and Bob in an extended nonlocal game, including 
\emph{unentangled strategies}, \emph{entangled strategies}
(or \emph{standard quantum strategies}), and \emph{commuting measurement
  strategies} \cite{JohnstonMRW16}.
(Additional classes of strategies, such as \emph{no-signaling strategies}, can
also be defined.)
In this paper we will only consider \emph{entangled strategies}, in which Alice
and Bob begin the game in possession of finite-dimensional quantum systems that
have been initialized as they choose.
They may then measure these systems in order to obtain answers to the referee's
questions.

In more precise terms, an entangled strategy for an extended nonlocal game,
specified by \mbox{$\pi:X\times Y \rightarrow [0,1]$} and 
$\{P_{a,b,x,y}\,:\,a\in A,\, b\in B,\, x\in X,\, y\in Y\} \subset \Pos(\R)$
as above, consists of these objects:
\begin{mylist}{8mm}
\item[1.]
  A state $\sigma \in \Density(\U\otimes\R\otimes\V)$, for $\U$ being the space
  corresponding to a register $\reg{U}$ held by Alice and $\V$ being the space
  corresponding to a register $\reg{V}$ held by Bob.
  This state represents Alice and Bob's initialization of the triple
  $(\reg{U},\reg{R},\reg{V})$ immediately before $\reg{R}$ is sent to the
  referee.
\item[2.]
  A measurement $\{A^x_a\,:\,a\in A\}\subset\Pos(\U)$ for each $x\in X$, 
  performed by Alice when she receives the question $x$, and a
  measurement $\{B^y_b\,:\,b\in B\}\subset\Pos(\V)$ for each $y\in Y$,
  performed by Bob when he receives the question $y$.
\end{mylist}
When Alice and Bob utilize such a strategy, their winning probability $p$
may be expressed as
\begin{equation}
  p = 
  \sum_{\substack{
      (x,y) \in X\times Y\\
      (a,b) \in A\times B\\
  }}
  \pi(x,y)\bigip{A^x_a \otimes P_{a,b,x,y} \otimes B^y_b}{\sigma}.
\end{equation}

The \emph{entangled value} of an extended nonlocal game represents the supremum
of the winning probabilities, taken over all entangled strategies.
If $H$ is the name assigned to an extended nonlocal game having a specification
as above, then we write $\omega^{\ast}_N(H)$ to denote the \emph{maximum}
winning probability taken over all entangled strategies for which
$\dim(\U\otimes\V) \leq N$, so that the entangled value of $H$ is
\begin{equation}
  \omega^{\ast}(H) = \lim_{N\rightarrow\infty}\omega^{\ast}_N(H).
\end{equation}

\subsection{Quantum-classical games}

A \emph{quantum-classical game} (or \emph{QC game}, for short) is specified
by the following objects:
\begin{mylist}{8mm}
\item[$\bullet$]
  A state $\rho \in \Density(\X\otimes\S\otimes\Y)$ of a triple of 
  registers $(\reg{X},\reg{S},\reg{Y})$.
\item[$\bullet$]
  A collection of measurement operators
  $\{Q_{a,b}\,:\,a \in A,\; b\in B\}\subset\Pos(\S)$, for finite and nonempty
  sets $A$ and~$B$.
\end{mylist}
From the referee's perspective, such a game is played as follows:
\begin{mylist}{8mm}
\item[1.]
  The referee prepares $(\reg{X},\reg{S},\reg{Y})$ in the state $\rho$, then
  sends $\reg{X}$ to Alice and $\reg{Y}$ to Bob.
\item[2.]
  Alice responds with $a\in A$ and Bob responds with $b\in B$.
\item[3.]
  The referee measures $\reg{S}$ with respect to the binary-valued
  measurement $\{Q_{a,b},\, \I-Q_{a,b}\}$.
  The outcome corresponding to the measurement operator $Q_{a,b}$ indicates
  that Alice and Bob \emph{win}, while the other measurement result indicates
  that they \emph{lose}.
\end{mylist}

Similar to extended nonlocal games, one may consider various classes of
strategies for QC games.
Again, we will consider only entangled strategies, in which Alice and Bob
begin the game in possession of finite-dimensional quantum systems initialized
as they choose.

More precisely, an entangled strategy for a QC game, specified by
$\rho \in \Density(\X\otimes\S\otimes\Y)$ and 
$\{Q_{a,b}\,:\,a \in A,\; b\in B\}\subset\Pos(\S)$ as above, consists of these
objects:
\begin{mylist}{8mm}
\item[1.]
  A state $\sigma\in\Density(\U\otimes\V)$, for $\U$ being the space
  corresponding to a register $\reg{U}$ held by Alice and $\V$ being the space
  corresponding to a register $\reg{V}$ held by Bob.
\item[2.]
  A measurement $\{A_a\,:\,a\in A\}\subset\Pos(\U\otimes\X)$ for Alice,
  performed on the pair $(\reg{U},\reg{X})$ after she receives $\reg{X}$ from
  the referee, and a measurement $\{B_b\,:\,b\in B\}\subset\Pos(\Y\otimes\V)$
  for Bob, performed on the pair $(\reg{Y},\reg{V})$ after he receives
  $\reg{Y}$ from the referee.
\end{mylist}
The winning probability of such a strategy may be expressed as
\begin{equation}
  p = \sum_{(a,b)\in A\times B}
  \bigip{A_a \otimes Q_{a,b} \otimes B_b}{W(\sigma \otimes \rho) W^{\ast}},
\end{equation}
where $W$ is the unitary operator that corresponds to the natural re-ordering
of registers consistent with each of the tensor product operators
$A_a \otimes Q_{a,b} \otimes B_b$ (i.e., the permutation 
$(\reg{U},\reg{V},\reg{X},\reg{S},\reg{Y})\mapsto
(\reg{U},\reg{X},\reg{S},\reg{Y},\reg{V})$).

%------------------------------------------------------------------------------%
\section{Construction and analysis}
\label{sec:construction}
%------------------------------------------------------------------------------%

In this section we will describe a construction of an extended nonlocal game
from any given QC game, and analyze the relationship between the constructed
extended nonlocal game and the original QC game.

\subsection{Construction}

Suppose that a QC game $G$, specified by a state 
$\rho \in \Density(\X\otimes\S\otimes\Y)$ and a collection of measurement
operators $\{Q_{a,b}\,:\,a \in A,\; b\in B\}\subset\Pos(\S)$, is given.
We construct an extended nonlocal game $H$ as follows:
\begin{mylist}{8mm}
\item[1.] Let $n = \dim(\X)$ and $m = \dim(\Y)$, let
  \begin{equation}
    X = \bigl\{1,\ldots,n^2\bigr\} 
    \quad\text{and}\quad 
    Y = \bigl\{1,\ldots,m^2\bigr\},
  \end{equation}
  and let $\pi:X\times Y\rightarrow[0,1]$ be the uniform probability
  distribution on these sets, so that $\pi(x,y) = n^{-2}m^{-2}$ for every
  $x\in X$ and $y\in Y$.

\item[2.]
  Let $\reg{R} = (\reg{X},\reg{Y})$, define
  \begin{equation}
    \xi = \tr_{\S}(\rho) \quad\text{and}\quad
    \xi_{a,b} = 
    \tr_{\S}\bigl[\bigl(\I_{\X}\otimes Q_{a,b} \otimes \I_{\Y}\bigr)\rho
      \bigr]
  \end{equation}
  for each $a\in A$ and $b\in B$, let
  \begin{equation}
    \bigl\{U_1,\ldots,U_{n^2}\bigr\}\subset\Unitary(\X)
    \quad\text{and}\quad
    \bigl\{V_1,\ldots,V_{m^2}\bigr\}\subset\Unitary(\Y)
  \end{equation}
  be orthogonal bases of unitary operators
  (such as the discrete Weyl operators, described in \cite{DattaFH06} for
  instance), and let
  \begin{equation}
    P_{a,b,x,y} = \I_{\X}\otimes\I_{\Y}
    - (U_x \otimes V_y)(\xi^{\t} - \xi_{a,b}^{\t})(U_x \otimes V_y)^{\ast}
  \end{equation}
  for every $a\in A$, $b\in B$, $x\in X$, and $y\in Y$.
\end{mylist}

One may observe that $P_{a,b,x,y}$ is indeed a measurement operator for each
$a\in A$, $b\in B$, $x\in X$, and $y\in Y$, meaning that
$0 \leq P_{a,b,x,y} \leq \I_{\X}\otimes\I_{\Y}$, by virtue of the fact that
$0 \leq \xi_{a,b} \leq \xi \leq \I$ for every $a\in A$ and $b\in B$.

The basic intuition behind this construction is as follows.
In the QC game $G$, the referee sends $\reg{X}$ to Alice and $\reg{Y}$ to Bob,
but in the extended nonlocal game $H$ it is Alice and Bob that give $\reg{X}$
and $\reg{Y}$ to the referee.
To simulate, within the game $H$, the sort of transmission that occurs in
$G$, it is natural to consider \emph{teleportation}---for if Alice provided the
referee with the register $\reg{X}$ in a state maximally entangled with a
register of her own, and Bob did likewise with $\reg{Y}$, then the referee
could effectively teleport a copy of $\reg{X}$ to Alice and a copy of $\reg{Y}$
to Bob.
Now, in an extended nonlocal game, the referee cannot actually perform
teleportation in this way: the question pair $(x,y)$ needs to be randomly
generated, independent of the state of the registers $(\reg{X},\reg{Y})$.
For this reason the game $H$ is based on a form of
\emph{post-selected teleportation}, where $x$ and $y$ are chosen randomly,
and then later compared with hypothetical measurement results that would
be obtained if the referee were to perform teleportation.
The details of the construction above result from a combination of this idea
together with algebraic simplifications.

\subsection{Game values}

It is not immediate that the construction above should necessarily translate
the basic properties of the game $G$ to the game $H$;
Alice and Bob are free to behave as they choose, which is not necessarily
consistent with the intuitive description of the game $H$ based on
teleportation suggested above.
An analysis does, however, reveal that the construction works as one would hope
(and perhaps expect).
In particular, we will prove two bounds on the value of the extended nonlocal
game $H$ constructed from a QC game $G$ as described above:
\begin{equation}
  \label{eq:two-bounds-omega-H}
  \omega^{\ast}_{nmN}(H) \geq 1 - \frac{1 - \omega^{\ast}_N(G)}{nm}
  \quad\text{and}\quad
  \omega^{\ast}_{N}(H) \leq 1 - \frac{1 - \omega^{\ast}_{nmN}(G)}{nm},
\end{equation}
for every positive integer $N$.
This implies that
\begin{equation}
  \omega^{\ast}(H) = 1 - \frac{1 - \omega^{\ast}(G)}{nm}.
\end{equation}
Moreover, $H$ inherits the same limiting behavior of $G$ with respect to
entangled strategies, meaning that if
$\omega^{\ast}_N(G) < \omega^{\ast}(G)$ for all $N\in\natural$, then
$\omega^{\ast}_N(H) < \omega^{\ast}(H)$ for all $N\in\natural$ as well.

We will begin with the first inequality in \eqref{eq:two-bounds-omega-H}.
Assume that an arbitrary strategy for Alice and Bob in the game $G$ is fixed:
Alice and Bob make use of a shared entangled state 
$\sigma \in \Density(\U\otimes\V)$, where $\dim(\U\otimes\V)\leq N$,
and their measurements are given by
\begin{equation}
  \{A_a\,:\,a\in A\}\subset\Pos(\U\otimes\X)
  \quad\text{and}\quad
  \{B_b\,:\,b\in B\}\subset\Pos(\Y\otimes\V),
\end{equation}
respectively.
The winning probability of this strategy in the game $G$ may be expressed as
\begin{equation}
  p = \sum_{(a,b)\in A\times B}
  \bigip{A_a \otimes Q_{a,b} \otimes B_b}{W (\sigma \otimes \rho) W^{\ast}},
\end{equation}
as was mentioned above, while the losing probability equals
\begin{equation}
  \label{eq:losing-in-G}
  q = \sum_{(a,b)\in A\times B}
  \bigip{A_a \otimes (\I - Q_{a,b})
    \otimes B_b}{W(\sigma \otimes \rho)W^{\ast}} = 1-p.
\end{equation}
We adapt this strategy to obtain one for $H$ as follows:
\begin{mylist}{8mm}
\item[1.]
  Alice will hold a register $\reg{X}'$, representing a copy of $\reg{X}$, and
  Bob will hold $\reg{Y}'$, representing a copy of $\reg{Y}$.
  The initial state of the register pairs $(\reg{X}',\reg{X})$ and
  $(\reg{Y}',\reg{Y})$ are to be the canonical maximally entangled states 
  \begin{equation}
    \ket{\psi} = \frac{1}{\sqrt{n}} \sum_{j = 1}^n \ket{j}\ket{j}
    \quad\text{and}\quad
    \ket{\phi} = \frac{1}{\sqrt{m}} \sum_{k = 1}^m \ket{k}\ket{k},
  \end{equation}
  respectively, where $n$ and $m$ are the dimensions of the spaces
  corresponding to the registers $\reg{X}$ and $\reg{Y}$.
  In addition, Alice holds the register $\reg{U}$ and Bob holds the register
  $\reg{V}$, with $(\reg{U},\reg{V})$ being prepared in the same shared
  entangled state $\sigma$ that is used in the strategy for $G$.

\item[2.]
  Upon receiving the question $x\in X$ from the referee, Alice performs the
  unitary operation $\overline{U_x}$ on $\reg{X}'$, then measures
  $(\reg{U},\reg{X}')$ with respect to the measurement $\{A_a\,:\,a\in A\}$
  to obtain an answer $a\in A$.
  Similarly, upon receiving $y\in Y$ from the referee, Bob performs
  $\overline{V_y}$ on $\reg{Y}'$, then measures $(\reg{Y}',\reg{V})$ with
  respect to $\{B_b\,:\,b\in B\}$ to obtain an answer $b\in B$.
\end{mylist}

The performance of this strategy can be analyzed by first ignoring the
specific initialization of the registers described in step 1, and defining a
measurement $\{R_0,R_1\}$ that determines, for an arbitrary initialization of
these registers, whether Alice and Bob win or lose by behaving as described in
step~2. 
In particular, the measurement $\{R_0,R_1\}$ is defined on the register tuple
$(\reg{U},\reg{X}',\reg{X},\reg{Y},\reg{Y}',\reg{V})$, the measurement operator
$R_0$ corresponds to a losing outcome, and $R_1$ corresponding to a winning
outcome.
These operators may be described as follows:
\begin{equation}
  \begin{aligned}
    R_0 & = 
    \frac{1}{n^2 m^2}
    \sum_{\substack{(x,y) \in X\times Y\\[0.3mm](a,b) \in A\times B}}
    (\I_{\U} \otimes U_x^{\t}) A_a (\I_{\U} \otimes \overline{U_x})
    \otimes (\I_{\X\otimes\Y} - P_{a,b,x,y})
    \otimes (V_y^{\t} \otimes \I_{\V}) B_b (\overline{V_y}\otimes \I_{\V})\\
    R_1 & = 
    \frac{1}{n^2 m^2}
    \sum_{\substack{(x,y) \in X\times Y\\[0.3mm](a,b) \in A\times B}}
    (\I_{\U} \otimes U_x^{\t}) A_a (\I_{\U} \otimes \overline{U_x})
    \otimes P_{a,b,x,y}
    \otimes (V_y^{\t} \otimes \I_{\V}) B_b (\overline{V_y}\otimes \I_{\V})
    = \I - R_0.
  \end{aligned}
\end{equation}

Now we may consider the initialization of the registers described in step~1.
For an arbitrary choice of operators $X\in\Lin(\U)$ and
$Y\in\Lin(\V)$ we have
\begin{equation}
  \bigip{R_0}{X \otimes \ket{\psi}\bra{\psi} \otimes \ket{\phi}\bra{\phi}
    \otimes Y}
  = \sum_{(a,b) \in A\times B}
    \bigip{
      A_a \otimes (\xi^{\t} - \xi^{\t}_{a,b}) \otimes B_b}{
      X \otimes \ket{\psi}\bra{\psi} \otimes 
      \ket{\phi}\bra{\phi}\otimes Y},
\end{equation}
by virtue of the fact that
$\bigl( \overline{U_x} \otimes U_x\bigr) \ket{\psi} = \ket{\psi}$
and $\bigl( \overline{V_y} \otimes V_y\bigr) \ket{\phi} = \ket{\phi}$
for every $x\in X$ and $y\in Y$.
Further simplifying this expression, one obtains
\begin{equation}
  \begin{aligned}
    \sum_{(a,b) \in A\times B}
    \bigip{
      A_a \otimes (\xi^{\t} - \xi^{\t}_{a,b}) \otimes B_b}{
      X \otimes \ket{\psi}\bra{\psi} \otimes 
      \ket{\phi}\bra{\phi}\otimes Y}\hspace{-5cm}\\
    & = \frac{1}{nm}\sum_{(a,b) \in A\times B}
    \bigip{A_a \otimes B_b}{X \otimes (\xi-\xi_{a,b}) \otimes Y}\\
    & = \frac{1}{nm}\sum_{(a,b) \in A\times B}
    \bigip{A_a \otimes (\I - Q_{a,b}) \otimes B_b}{
      X \otimes \rho \otimes Y}.
  \end{aligned}
\end{equation}
By expressing the initial state $\sigma$ of $(\reg{U},\reg{V})$ as
$\sigma = \sum_i X_i \otimes Y_i$ and making use of the bilinearity of
the above expression in $X$ and $Y$, one finds that the losing
probability of Alice and Bob's strategy for $H$ is equal to
$q/(nm)$, for $q$ being the losing probability \eqref{eq:losing-in-G} for their
original strategy for $G$.

Optimizing over all strategies for $G$ that make use of an initial shared state
having total dimension at most $N$ yields the required inequality
\begin{equation}
  \omega^{\ast}_{nmN}(H) \geq 1 - \frac{1 - \omega^{\ast}_N(G)}{nm}.
\end{equation}

Next we will prove the second inequality in \eqref{eq:two-bounds-omega-H}.
Assume that an arbitrary strategy for Alice and Bob in the extended nonlocal
game $H$ constructed from $G$ is fixed: the strategy consists of an initial
state $\sigma \in \Density(\U\otimes(\X\otimes\Y)\otimes\V)$ for the registers
$(\reg{U},(\reg{X},\reg{Y}),\reg{V})$, where $\dim(\U\otimes\V)\leq N$, along
with measurements
\begin{equation}
  \bigl\{A^x_a\,:\,a\in A\bigr\}\subset\Pos(\U)
  \quad\text{and}\quad
  \bigl\{B^y_b\,:\,b\in B\bigr\}\subset\Pos(\V)
\end{equation}
for Alice and Bob, respectively, for each $x\in X$ and $y \in Y$.
The winning probability of this strategy may be expressed as
\begin{equation}
  p = \frac{1}{n^2 m^2}
  \sum_{\substack{
      (x,y) \in X\times Y\\
      (a,b) \in A\times B\\
  }}
  \Bigip{
    A^x_a \otimes P_{a,b,x,y} \otimes B^y_b}{\sigma}
\end{equation}
while the losing probability is
\begin{equation}
  \label{eq:losing-in-H}
  q = \frac{1}{n^2 m^2}
  \sum_{\substack{
      (x,y) \in X\times Y\\
      (a,b) \in A\times B\\
  }}
  \Bigip{
    A^x_a \otimes (\I - P_{a,b,x,y}) \otimes B^y_b}{\sigma} = 1-p.
\end{equation}
We adapt this strategy to give one for $G$ as follows:
\begin{mylist}{8mm}
\item[1.]
  Let $\reg{X}'$ and $\reg{Y}'$ represent copies of the registers $\reg{X}$ and
  $\reg{Y}$.
  Alice and Bob will initially share the registers
  $(\reg{U},\reg{X}',\reg{Y}',\reg{V})$ initialized to the state 
  $\overline{\sigma}$, with Alice holding $(\reg{U},\reg{X}')$ and Bob holding
  $(\reg{Y}',\reg{V})$.
\item[2.]
  Upon receiving $\reg{X}$ from the referee, Alice first measures the pair
  $(\reg{X}',\reg{X})$ with respect to the basis
  $\bigl\{ (\I \otimes U_x^{\ast})\ket{\psi}\,:\,x\in X\bigr\}$.
  For whichever outcome $x\in X$ she obtains, she then measures $\reg{U}$ with
  respect to the measurement
  \begin{equation}
    \Bigl\{\overline{A^x_a}\,:\,a\in A\Bigr\} \subset \Pos(\U)
  \end{equation}
  to obtain an outcome $a\in A$.
  Bob does likewise, first measuring $(\reg{Y}',\reg{Y})$ with respect to the
  basis $\bigl\{ (\I \otimes V_y^{\ast})\ket{\phi}\,:\,y\in Y\bigr\}$,
  and then measuring $\reg{V}$ with respect to the measurement
  \begin{equation}
    \Bigl\{\overline{B^y_b}\,:\,b\in B\Bigr\} \subset \Pos(\V)
  \end{equation}
  for whichever outcome $y\in Y$ is obtained.
\end{mylist}

Now let us consider the probability with which this strategy wins in $G$.
The state of the registers
$(\reg{U},\reg{X}',\reg{X},\reg{S},\reg{Y},\reg{Y}',\reg{V})$ 
immediately after the referee sends $\reg{X}$ to Alice and $\reg{Y}$ to Bob is
given by
\begin{equation}
  W ( \overline{\sigma} \otimes \rho) W^{\ast},
\end{equation}
where $W$ is a unitary operator that corresponds to a permutation of
registers:
\begin{equation}
  (\reg{U},\reg{X}',\reg{Y}',\reg{V},\reg{X},\reg{S},\reg{Y}) \mapsto
  (\reg{U},\reg{X}',\reg{X},\reg{S},\reg{Y},\reg{Y}',\reg{V}).
\end{equation}
We may define a measurement $\{R_0,R_1\}$ on the register tuple
$(\reg{U},\reg{X}',\reg{X},\reg{S},\reg{Y},\reg{Y}',\reg{V})$
representing the outcome of the game, with $R_0$ corresponding to a losing
outcome and $R_1$ corresponding to a winning outcome.
We have
\begin{equation}
  \begin{aligned}
    R_0 & =
    \sum_{\substack{(x,y) \in X\times Y\\[0.3mm](a,b) \in A\times B}}
    \overline{A^x_a} \otimes 
    (\I \otimes U_x^{\ast})\ket{\psi}\bra{\psi}(\I \otimes U_x)
    \otimes (\I - Q_{a,b}) \otimes
    (V_y^{\ast}\otimes \I)\ket{\phi}\bra{\phi}(V_y\otimes\I)
    \otimes \overline{B^y_b}\\
    R_1 & =
    \sum_{\substack{(x,y) \in X\times Y\\[0.3mm](a,b) \in A\times B}}
    \overline{A^x_a} \otimes 
    (\I \otimes U_x^{\ast})\ket{\psi}\bra{\psi}(\I \otimes U_x)
    \otimes Q_{a,b} \otimes
    (V_y^{\ast}\otimes\I)\ket{\phi}\bra{\phi}(V_y\otimes\I)
    \otimes \overline{B^y_b}.
  \end{aligned}
\end{equation}
Simplifying expressions for the probability that Alice and Bob lose yields
\begin{equation}
  \bigip{R_0}{W ( \overline{\sigma} \otimes \rho) W^{\ast}}
  = \frac{1}{nm}
  \sum_{\substack{
      (x,y) \in X\times Y\\
      (a,b) \in A\times B\\
  }}
  \Bigip{A^x_a \otimes (\I - P_{a,b,x,y}) \otimes B^y_b}{\sigma}
  = nm q,
\end{equation}
for $q$ being the losing probability \eqref{eq:losing-in-H} for their original
strategy for $H$.

Optimizing over all strategies for $H$ that make use of an initial shared state
for which Alice and Bob's total dimension is at most $N$ yields the inequality
\begin{equation}
  \omega^{\ast}_{N}(H) \leq 1 - \frac{1 - \omega^{\ast}_{nmN}(G)}{nm}.
\end{equation}

%-----------------------------------------------------------------------------%
\section{Discussion}
\label{sec:discussion}
%-----------------------------------------------------------------------------%

As was mentioned in the introduction, Regev and Vidick \cite{RegevV15} have
identified examples of QC games for which Alice and Bob can never achieve
optimality by using a finite-dimensional entangled strategy.
To be more precise, they prove that there exists a QC game\footnote{
  Their games fall into a category of QC games that they call
  \emph{quantum XOR games}, in which $A = B = \{0,1\}$ and only the parity
  $a\oplus b$ of Alice and Bob's answers is relevant to the referee's
  determination of whether they win or lose.}
$G$ (and in fact a family of such games) for which it holds that
$\omega^{\ast}_N(G) < 1$ for all $N\in\natural$, while $\omega^{\ast}(G) = 1$.
By applying our construction to any such game, we obtain an extended nonlocal
game $H$ with the property that $\omega^{\ast}_N(H) < 1$ for all
$N\in\natural$, while $\omega^{\ast}(H) = 1$.

In greater detail, by taking the simplest known example of a QC game $G$ with
the property just described, and applying our construction (along with minor
simplifications), one obtains an extended nonlocal game as follows:
\begin{mylist}{8mm}
\item[1.] Let $\X = \Y = \complex^3$ and let $U_1,\ldots,U_9$ be the
  discrete Weyl operators acting on $\complex^3$.
  Also define
  \begin{equation}
    \begin{aligned}
      \ket{\gamma_0} & = \frac{1}{\sqrt{2}}\ket{0}\ket{0}
      + \frac{1}{2}\ket{1}\ket{1} + \frac{1}{2}\ket{2}\ket{2},\\
      \ket{\gamma_1} & = \frac{1}{\sqrt{2}}\ket{0}\ket{0}
      - \frac{1}{2}\ket{1}\ket{1} - \frac{1}{2}\ket{2}\ket{2}.
    \end{aligned}
  \end{equation}
\item[2.]
  Alice and Bob give a pair of registers $(\reg{X},\reg{Y})$ to the referee,
  initialized as they choose.
  The referee randomly chooses $x,y\in\{1,\ldots,9\}$ uniformly and
  independently at random, then sends $x$ to Alice and $y$ to Bob.
  Alice and Bob respond with binary values $a,b\in\{0,1\}$, respectively.
\item[3.]
  The referee computes $c = a\oplus b$, then measures the pair
  $(\reg{X},\reg{Y})$ with respect to the measurement
  \begin{equation}
    \bigl\{
    \I_{\X}\otimes\I_{\Y} - (U_x \otimes U_y)\ket{\gamma_c}\bra{\gamma_c}(U_x
    \otimes U_y)^{\ast},\;
    (U_x \otimes U_y)\ket{\gamma_c}\bra{\gamma_c}(U_x \otimes
    U_y)^{\ast}\bigr\}.
  \end{equation}
  The first outcome represents a win for Alice and Bob, and the second a loss.
  (Note that here we have scaled the losing measurement operator by a factor of
  two in comparison to what is described in the construction, which has the
  effect of doubling the losing probability for every strategy of Alice and
  Bob.)
\end{mylist}
Assuming Alice and Bob initially entangle the pair $(\reg{X},\reg{Y})$ with
finite-dimensional registers of their own, they can never win the game with
certainty, but they can approach certainty by using increasingly large
systems.

%-----------------------------------------------------------------------------%
\subsection*{Acknowledgments}
%-----------------------------------------------------------------------------%

This work was partially funded by Canada's NSERC.
We thank Marco Piani, William Slofstra, and Thomas Vidick for helpful
comments and discussions.

\bibliographystyle{alpha}
\bibliography{games}

\end{document}